\documentclass{elsart}
\usepackage{amsmath}
\usepackage[english]{babel}
\usepackage{graphicx}

\begin{document}
\begin{frontmatter}

\title{THEORY OF SUB-10 FS GENERATION IN KERR-LENS MODE-LOCKED SOLID-STATE LASERS
WITH A COHERENT SEMICONDUCTOR ABSORBER}

\author{V. L. Kalashnikov\thanksref{l1}}
\thanks[l1]{The author is a gratefull to Dr. I. G. Poloyko and D. O. Krimer for the helpful discussions and suggestions}

\address{International Laser Center, 65 Skorina Ave., Bldg. 17, Minsk, 220027 Belarus tel./fax: /375-0172/ 326-286}
\ead{vkal@ilc.unibel.by}
\ead[url]{http://www.geocities.com/optomaplev}

\begin{abstract}
The results of the study of ultra-short pulse generation in continuous-wave
Kerr-lens mode-locked (KLM) solid-state lasers with semiconductor saturable
absorbers are presented. The issues of extremely short pulse generation are
addressed in the frames of the theory that accounts for the coherent nature
of the absorber-pulse interaction. We developed an analytical model that bases
on the coupled generalized Landau-Ginzburg laser equation and Bloch equations
for a coherent absorber. We showed, that in the absence of KLM semiconductor
absorber produces 2\( \pi  \) - non-\textit{sech-}pulses of self-induced transparency,
while the KLM provides an extremely short \textit{sech}-shaped pulse generation.
2\( \pi  \)- and \( \pi  \) - \textit{sech}-shaped solutions and variable-area
chirped pulses have been found. It was shown, that the presence of KLM removes the
limitation on the minimal modulation depth in absorber. An automudulational
stability and self-starting ability were analyzed, too.
\end{abstract}


\end{frontmatter}

\section{Introduction}
\noindent
A recent progress in ultra-fast lasers has resulted in the generation of sub-10
fs pulses, which is close to the fundamental limit defined by the light wave
period in visible and near-IR \cite{Keller1}. Today a basic technique for fs-generation
is the Kerr-lens mode locking (KLM) \cite{Haus1} in the combination with a slow saturation
of interband or excitonic transitions in semiconductor structure \cite{Keller2}. It
was shown \cite{Keller3}, that in both cases a quasi-soliton formation plays a key
role in the stabilization of ultra-short pulse generation down to shortest possible
pulse durations. Moreover, the quasi-soliton generation due to its highest stability
and simple pulse form is the subject for the use of analytical and semi-analytical
approaches, which are of highest theoretical and practical interest that stimulates
the investigations in this field.

As was shown in \cite{My1,My2}, the strong nonlinear effects in semiconductor absorbers,
such as absorption linewidth enhancement and Stark effect, transform the laser
dynamics essentially, that causes an additional pulse stabilization and compression.
Since the pulse durations in modern ultrafast lasers are comparable or shorter
than the absorber dephasing time \( t_{coh} \), the coherent nature of the
pulse-semiconductor interaction has to be taken into consideration. A coherent
absorber mode locking has been analyzed in refs. \cite{Kozlov1,Kozlov2,Komarov,Kalosha}. In particular,
it was shown in \cite{Kozlov1,Kozlov2,Komarov} that the dynamical gain saturation is essential for
the coherent quasi-soliton generation in the lasers. However, the dynamical gain
saturation is negligible in femtosecond solid-state lasers, where the dominating
nonlinear factors are the self-phase modulation (SPM) and the self-focusing.
Numerical simulations \cite{Kalosha} have demonstrated the generation of fs-pulses
of self-induced transparency in solid-state laser with semiconductor quantum-well
absorber in the absence of KLM. However, as was shown in \cite{Chr},
the self-focusing is essential in femtosecond time domain and should be taken into
account in the analysis. Furthermore, an analytical approach to this problem
can provide a new insight into the physics of fs-lasers and semiconductor optical
devices.

Here we present a study of fs-pulse generation in cw solid-state KLM - laser with semiconductor absorber. We developed an analytical model that accounts
for a coherent pulse-semiconductor interaction, SPM, KLM, group-velocity dispersion
(GVD) and gain saturation by the full pulse energy. The condition of the \textit{sech}-shaped
pulse formation was found and the contribution of Kerr-lensing to pulse characteristics
was considered. We showed that the generation of chirp-free \( 2\pi  \)- and
\( \pi  \)- pulses as well as chirped pulses with the variable area is possible.
The obtained solutions are stable against laser noise and automodulational instability.

\section{Model}

Based on the slowly varying envelope approximation for the field amplitude \( a(t) \),
let consider a distributed system including saturable gain, linear loss and
SPM with coefficients \( \alpha  \), \( \gamma  \) and \( \beta  \), respectively,
Kerr-lens-induced fast saturable absorption with saturation intensity \( 1/\sigma  \),
GVD with coefficient \( D \) and bandwidth limiting element with transmission
bandwidth \( 1/t_{f} \), where \( t_{f} \) defines the minimal pulse duration
and is equal to 2.5 fs for Ti: sapphire laser. For the sake of staying in framework
of analytical approach we used a two-level model for quantum-well semiconductor
absorber. This assumption is valid for quantum-confined semiconductor structures
utilized in mode-locked fs-lasers (see, for example, \cite{Keller3,Kalosha}; last work
contains some generalizations of model in framework of numerical analysis).

When the pulse duration \( t_{p} \) is much shorter than the dephasing time
in absorber \( t_{coh} \) and the field intensity \( |a(t)|^{2} \)
is not enough for the Stark effect manifestation, the pulse-semiconductor interaction
obeys the Bloch equations \cite{Allen}:

\begin{equation}
\label{e1}
\frac{du}{dt}=(\Delta -\frac{d\phi }{dt})v+qaw,\qquad \frac{dv}{dt}=-(\Delta -\frac{d\phi }{dt})u,\qquad \frac{dw}{dt}=-qau,
\end{equation}

\noindent where \( u(t) \), \( v(t) \) and \( w(t) \) are the slowly varying
envelopes of the polarization quadrature components and the population difference,
respectively, \( q=d/\hbar  \), \( d=0.28\cdot e \) {[}Coulomb\( \times  \)nm{]}
is the dipole momentum, \( e \) is the elementary charge (in our calculations
we used the parameters of GaAs/AlAs absorber, the saturation energy \( E_{a}=50\mu J/cm^{2} \)
and \( t_{coh}=50\; fs \)), \( \Delta  \) is the mismatch between optical
resonance and pulse carrier frequency, \( \phi  \) is the instant field phase.
An initial saturable absorption \( \gamma _{a}=2\pi \cdot N\cdot d\cdot 2\omega \cdot z_{a}\cdot t_{coh}/(c\hbar )=0.01 \)
(\( \omega  \) is the field frequency) corresponds to the carriers density
\( N=\gamma _{a}E_{a}/(\hbar \cdot \omega \cdot z_{a})=2\cdot 10^{18}\; cm^{-3} \)
and the thickness of semiconductor absorber \( z_{a}=10\; nm \).

The laser part of the master equation is the generalized Landau-Ginzburg equation
\cite{Haus1}. Then the master equation can be written as:

\begin{eqnarray}
\label{e2}
\frac{\partial a(z,t)}{\partial z}=\left[ \alpha -\gamma +i\theta +\delta \frac{\partial }{\partial t}+(t_{f}^{2}+iD)\frac{\partial ^{2}}{\partial t^{2}}+\frac{\sigma -i\beta }{\eta ^{2}}\left| a\right| ^{2}\right] a+ &  & \nonumber \\
\left[ \frac{2\pi Nz_{a}\omega d}{c}u-\frac{2\pi Nz_{a}d}{c}\frac{dv}{dt}\right] ,
\end{eqnarray}

\noindent where \( z \) is the longitudinal coordinate normalized to the cavity
length, i.e. the number of the cavity round-trip, \( c \) is the light velocity,
\( \theta  \) and \( \delta  \) are the phase and time field delays of the
field after the cavity round-trip, respectively. We neglected the spatial effects
in the absorber. Later we will consider only steady-state pulse-like solutions,
which allows to eliminate the dependence on \( z \). Let normalize the times
to \( t_{f} \) and the field to \( q\cdot t_{f} \) (note, however, that we
will use a dimensional pulse duration in Figs. \ref{f2} and \ref{f4}). Then \( \beta  \) and
\( \sigma  \) are normalized to \( 2(q\cdot t_{f})^{2}/(n\cdot c\cdot \varepsilon _{0})=5\cdot 10^{-12}\; cm^{2}/W \),
where \( n \) is the index of refractivity, \( \varepsilon _{0} \) is the
permittivity and \( t_{f}=2.5 \) \( fs \) (Ti: sapphire laser). With these
normalizations, \( \sigma =0.14 \) corresponds to the saturation parameter
of Kerr-lens induced fast saturable absorber of \( 10^{7}\; W \) and \( 30\; \mu m \)
spot size in active medium. Dimensionless SPM parameter \( \beta  \) is equal
to \( 0.26 \) for \( 1\; mm \) Ti: sapphire crystal. Additionally, we introduce
an important control parameter \( \eta  \), which is governed by 1) the ratio
between the size of generation mode in active medium and in semiconductor absorber
or by 2) the reflectivity of the upper surface of the semiconductor saturable
device. Formally, the variation of \( \eta  \) means the variation of relative
contribution of SPM, Kerr-lens-induced saturable absorption and saturable gain
with respect to the saturable absorption in semiconductor.

As we will see later, the gain saturation by the full pulse energy is an important
factor for the pulse stability and thus should be taken into account. The simplest
way to do this is to use a quasi-two level model for active medium. After some
calculations for the gain saturated by the full pulse energy \( E \) in the
steady-state condition we have \( \alpha =\frac{P\alpha _{max}}{P+\tau E/\eta ^{2}+1/T_{r}} \)
, where \( \alpha _{max} \) is the gain for the full population inversion,
\( T_{r} \) is the gain relaxation time normalized to the cavity period \( T_{cav} \),
\( P=\sigma _{14}T_{cav}I_{p}/(h\cdot \nu _{p}) \) is the dimensionless pump
intensity, \( \nu _{p} \) is the pump frequency, \( \sigma _{14} \) is the
absorption cross section of active medium, \( I_{p} \) is the pump intensity,
\( \tau =6.25\cdot 10^{-4} \) is the normalized inverse energy of the gain
saturation.

Later we will consider the different realizations of our model aimed to investigation
of the soliton-like solution of the system (\ref{e1}, \ref{e2}).

\section{Pulse of the self-induced transparency in the absence of KLM}

In the beginning we consider the case of chirp-free pulse-like solutions. After
integration of the equations (\ref{e1}), the master equation (\ref{e2}) reads as:

\begin{eqnarray}
\label{e3}
\frac{\partial a(z,t)}{\partial z}=\left[ \alpha -\gamma +i\theta +\delta \frac{\partial }{\partial t}+(1+iD)\frac{\partial ^{2}}{\partial t^{2}}+\frac{\sigma -i\beta }{\eta ^{2}}\left| a\right| ^{2}\right] a- &  & \nonumber \\
\frac{\gamma _{a}}{t_{coh}}\sin (\psi (z,t)),
\end{eqnarray}

\noindent where \( \psi (z,t)= \) \( \int\limits _{-\infty }^{t}a(z,t\prime )dt\prime  \)
is the pulse area (note, that the field and time are the dimensionless quantities
here). Under steady-state condition (the pulse envelope is independent on \textit{z}),
an integro-differential Eq. (\ref{e3}) results in differential equation

\begin{eqnarray}
\label{e4}
\left[ (\alpha -\gamma +i\theta )\frac{d}{dt}+\delta \frac{d^{2}}{dt^{2}}+(1+iD)\frac{d^{3}}{dt^{3}}+\frac{\sigma -i\beta }{\eta ^{2}}\left( \frac{d\psi (t)}{dt}\right) ^{2}\frac{d}{dt}\right] \psi (t)- &  & \nonumber \\
\frac{\gamma _{a}}{t_{coh}}\sin (\psi (t))=0.
\end{eqnarray}

\noindent In the absence of the lasing factors such as linear loss, frequency
filtering, SPM and GVD we have a well-known nonlinear equation with \( 2\pi  \)-soliton solution in the form \( a(t)=a_{0}sech(t/t_{p}) \),
where \( a_{0} \) is the amplitude, \( t_{p} \) is the duration \cite{Allen}.
But this solution does not satisfy the full Eq. (\ref{e4}) in the absence of KLM (\( \sigma =0 \)). 

Now let consider the Eq. (\ref{e4}) without KLM, SPM and GVD (\( \sigma =\beta =D=\theta =0 \)).
The substitution \( \psi (t)=x \), \( d\psi (t)/dt=y(x) \) (``area-amplitude''
representation) reduces the third-order Eq. (\ref{e4}) to the second-order one:

\begin{equation}
\label{e5}
\left[ \left( \frac{d^{2}y}{dx^{2}}\right) y+\left( \frac{dy}{dx}\right) ^{2}+\delta \frac{dy}{dx}+(\alpha -\gamma )\right] y-\frac{\gamma _{a}}{t_{coh}}\sin (x)=0.
\end{equation}

\noindent We solved this equation numerically and found a non-\textit{sech}-shaped
\( 2\pi  \)-solutions for it (see Fig. \ref{f1}, where numerical solution is shown
in comparison with the \textit{sech}-shaped pulse). This clearly indicates on
some additional nonlinear factors which are necessary for the \textit{sech}-pulse
formation and which are lacking in Eq. (\ref{e5}).

However, \( 2\pi  \)- non-\textit{sech} solution needs a more detailed investigation
because of it may be relevant mechanism for extremely short pulse generation
in real lasers. As is known \cite{Keller3}, the main mechanism of destabilization of
fs-pulses is the noise generation as result of loss saturation in absorber.
In the case of \( 2\pi  \)-pulse formation, a Rabi flopping of the absorber
population suppresses the noise behind the pulse tail, that stabilizes fs-generation
\cite{Kalosha}.

To investigate the pulse-like solutions of Eq. (\ref{e5}) analytically, we used a harmonic
approximation: \( y(x)=a_{1}sin(x/2)+\sum\limits_{m > 2} {a_m \sin (\frac{{m\pi }}{2})} \),
which is usual for analysis of oscillating processes (see, for example, \cite{Ros}).
I our case this approximation is most appropriate for ``area-amplitude'' representation
because of that allows to describe the full pulse area but not only behavior
of the field amplitude in the vicinity of pulse maximum as in the case of polynomial
expansion. We select the subharmonic corresponding to \( 2\pi  \)-pulse generation
(first term) and ``high frequency'' corrections to it. Retaining only the
first term, in the ``area-amplitude'' representation, we arrive to the solution
\( a_{1}= \) \( 2\sqrt{2(\alpha -\gamma )} \), \( \delta =\gamma _{a}/(2(\alpha -\gamma )t_{coh}) \),
\( t_{p}=2/a_{1} \), which corresponds to \textit{sech}-shaped solution in
the ``time-amplitude'' representation. The relations between pulse parameters
are analogues to that ones for \( 2\pi  \)-\textit{sech}-shaped solution, but
an additional relation between pulse amplitude and dissipative coefficients
\( \alpha  \) and \( \gamma  \) appears.

Fig. \ref{f2} (curves 1 and 2) presents the pulse durations for two physical approximated
solutions of Eq. (\ref{e5}). One can see, that the coherent absorber provides sub-10
fs pulse generation starting from some minimal pump. An important feature of
the obtained solution is the positive difference between saturated gain and linear loss
coefficients \( \alpha -\gamma >0 \), which imposes an important requirement
on the possible minimal saturable loss coefficient \( \gamma _{a} \), which
is necessary for stable pulse generation. So, there is the minimal modulation
depth of absorber that confines the region of the pulse stability against laser
noise. As it was shown in \cite{Haus1}, the pulse is stable if the net-gain outside
pulse is negative that results in the condition \( \alpha -\gamma -\gamma _{a}<0 \),
i.e. \( \gamma _{a}>\alpha -\gamma  \). The dependence of the minimal modulation
depth on the pump is shown in Fig. \ref{f3} for two depicted in Fig. \ref{f2} solutions of Eq. (\ref{e5}) (lower curve corresponds to the solution with larger
duration, the dashed curve depicts the generation threshold, hatching shows
a corresponding stability zone). As one can see, the pulse stabilization against
laser noise is possible only for the solution with longer duration. The stability
range widens in \( \gamma _{a} \), however at the cost of pump growth. It should
be noted, that in the absence of a coherent mechanism the pulse stabilization
can be caused only by essential contribution of dynamical gain saturation \cite{Haus2,My3}, that is not possible in fs-time domain, or by ``soliton mode locking''
\cite{Keller3} due to SPM and GVD balance. Self-induced transparency in semiconductor
absorber produces a quite different mechanism of pulse stabilization, which
does not involve any additional nonlinear processes.

Now let introduce into equations the SPM and GVD terms, that corresponds
to the real-world femtosecond lasers. Assuming a chirp-free (i.e. pure real) nature
of possible solution we can reduce Eq. (\ref{e4}) to the first-order equation:

\begin{equation}
\label{e6}
\left[ \delta \frac{dy(x)}{dx}+\frac{\beta }{\eta ^{2}D}y(x)^{2}+(\alpha -\gamma -\frac{\theta }{D})\right] y(x)-\frac{\gamma _{a}}{t_{coh}}\sin x=0.
\end{equation}

\noindent With the above described harmonic approximation we have the following
solutions: \( \theta =3\beta \cdot a_{1}^{2}/(4\eta ^{2})+D(\alpha -\gamma ) \),
\( \delta =4\gamma _{a}/(t_{coh}\cdot a_{1}^{2}) \), \( a_{1}=2\sqrt{3(\alpha -\gamma )} \),
\( t_{p}=2/a_{1} \) for \( 2\pi  \)-pulse. The pulse durations are presented
in Fig. \ref{f2} by curves 1' and 2'. The stable solution has a slightly
longer duration than the unstable one.

Formally, the obtained solution is the solution of the laser part of master
equation that in the same time satisfies the Bloch equations. The stability
of the solution results from the self-induced transparency in absorber, when
the pulse propagates in the condition of the positive net-gain but noise is
suppressed due to Rabi flopping of the absorber population. We do not analyze
the automodulational stability \cite{My4} of the solution since it has been demonstrated
directly by the numerical simulation in \cite{Kalosha}.

Thus we can conclude, that there is not the generation of coherent \textit{sech}-shaped
pulse in the absence of Kerr-lens-induced fast saturable absorption. But the
generation of non-\textit{sech} \( 2\pi  \)-pulse takes place, which imposes
a limitation on minimal modulation depth of the absorber with subsequent growth
of generation threshold. Nevertheless, as result of the coherent \( 2\pi  \)-pulse
formation, the pulse stabilization in this case is possible without contribution
of dynamical gain saturation, or SPM and GVD. In the next section we take into
account the contribution of KLM.

\section{Coherent \protect\( 2\pi \protect \)-\textit{sech}-pulse in the presence of
KLM}

The presence of KLM is described by the term \( \sigma |a|^{2} \) in Eq. (\ref{e3}).
In this case there is an exact \( 2\pi  \)-\textit{sech}-pulse solution of Eq.
(\ref{e3}), however for a strict relation between \( \sigma  \) and \( \eta  \),
so that \( \sigma =\frac{\eta ^{2}}{2} \) . The \textit{sech}-shaped solution
has the following parameters:

\begin{equation}
\label{e7}
a_{0}=\frac{2}{t_{p}},\qquad t_{_{p}}=\frac{1}{\sqrt{\gamma -\alpha }},\qquad \delta =\frac{\gamma _{a}}{t_{coh}(\gamma -\alpha )}.
\end{equation}

\noindent There are two distinct features of \( 2\pi  \)-\textit{sech}-pulse
generation: 1) the condition \( \alpha -\gamma <0 \) is satisfied automatically
and, consequently, there is no limitation on the minimal modulation depth of
the absorber; 2) the expression for the pulse duration is precisely same as
for the case of pure fast saturable absorber mode locking \cite{Haus1}, that suggests
that the KLM is the main mechanism determining the pulse duration. This conclusion
corroborates with the results of ref. \cite{Chr}. The action of the coherent absorber
determines the pulse delay \( \delta  \) and imposes a restriction on the pulse
area, i. e. the relation between pulse duration and amplitude. Pulse duration
in the presence of KLM is shown in Fig. \ref{f2} by dotted curve 3. As is seen, the
pulse duration is much shorter than for the case with no KLM, especially for
the small pump. As the contribution of semiconductor absorber is increased (\( \eta  \)
approaches 1, curve 1 in Fig. \ref{f4}) the pulse duration is reduced down to the limit
of the validity of the slowly varying envelope approximation.

An explanation of the additional relation between the parameters of Kerr-lens-induced
saturable absorber and semiconductor absorber is as follows: a \textit{sech}-shaped
solution satisfies both pure laser equation and the Bloch equations, but the
coherent interaction discriminates the special cases of \( n\pi  \)-area pulses,
in particular \( 2\pi  \)-pulses, which are provided by \( \sigma =\eta ^{2}/2 \)
relation.

Let us study an automodulational stability of the coherent laser pulse, which
we showed before to be very important factor in fs-lasers \cite{My4}. We used
an aberrationless approximation, which assumes an approximately unchanged form
of solution and \( z \)-dependence for the pulse parameters. The substitution
of the pulse envelope in Eq. (\ref{e3}) with following expansion into the time series
yields:

\begin{eqnarray}
\label{e8}
\frac{da_{0}}{dz}=2\frac{(\alpha -\gamma )\eta ^{2}t_{p}^{2}-\eta ^{2}+4\sigma }{\eta ^{2}t_{p}^{3}},\qquad \frac{dt_{p}}{dz}=4\frac{\eta ^{2}-2\sigma }{a_{0}\eta ^{2}t_{p}^{2}}, &  & \nonumber \\
\theta =\frac{2(D+4\frac{\beta }{\eta ^{2}})}{a_{0}t_{p}^{3}},\qquad \delta =2\frac{\gamma _{a}t_{p}}{t_{coh}a_{0}}. & 
\end{eqnarray}

\noindent Eqs. (\ref{e8}) were derived for a chirp-free solution with the dispersion
\( D=-2\beta /\eta ^{2} \) exactly compensating SPM. To be self-consistent
the system (\ref{e8}) should be completed by the additional relation between pulse
duration and amplitude arising from the Bloch equations. After some calculations
we have the explicit expressions, which determine the pulse stability. The pulse
is stable if the Jacobian of the right-hand sides of first two Eqs. (\ref{e8}) has
only non-positive real parts of eigenvalues. The condition for amplitude perturbation
decay \( -4(\gamma -\alpha )^{2}<0 \) is satisfied automatically. For \( \sigma =\eta ^{2}/2 \)
the pulse possesses a marginal stability with respect to the evolution of pulse
duration. However for any \( \sigma <\eta ^{2}/2 \) the pulse stability condition
with respect to duration evolution is satisfied.

Thus, we have analyzed the characteristics of \( 2\pi  \)-\textit{sech}-pulses
generated in KLM-laser with semiconductor coherent absorber. The main features
here are: the ``automatic'' stabilization against laser noise and against
pulse automodulations, and, also, the pulse duration decrease down to sub-10
fs. However, equations describing this physical situation allow yet another
type of analytical solutions.

\section{Coherent \protect\( \pi \protect \)-\textit{sech}-pulse and chirped pulses
in the presence of KLM}

As one can see from the previous part of our work, the ultrashort pulse is the
soliton-like solution for the both laser part of the master equation and the
Bloch equations. Now we will consider the complex anzatz \( a(t)=a_{0}sech(\frac{t}{t_{p}})^{1-i\zeta } \)
describing a pulse with chirp \( \zeta  \). It is known \cite{Allen}, that the
Eqs. (\ref{e1}) have a \textit{sech}-shaped solutions in form of chirp-free \( \pi  \)-pulse
or \textit{sech}-shaped chirped solution, when the following relations hold:

\( u(t)=u_{0}sech(\frac{t}{t_{p}}), \ v(t)=v_{0}sech(\frac{t}{t_{p}}), \ w(t)=\tanh (\frac{t}{t_{p}}), \ \frac{d\phi (t)}{dt}=\frac{\zeta }{t_{p}}\tanh (\frac{t}{t_{p}}), \)

\noindent where

\( a_{0}=\frac{\sqrt{1+\zeta ^{2}}}{t_{p}},\quad u_{0}=-\frac{1}{\sqrt{1+\zeta ^{2}}},\quad v_{0}=\frac{\zeta }{\sqrt{1+\zeta ^{2}}}. \)

\noindent A chirp-free solution is a \( \pi  \)-pulse, which is obviously unstable
in the absorber since the full population inversion behind the pulse tail amplifies
the noise. However, another nonlinear factors in KLM-laser can stabilize the
pulse and this requires a corresponding consideration.

Parameters of the chirp-free \( \pi  \)-pulse in KLM-laser are:

\begin{equation}
\label{e9}
a_{0}=\frac{1}{t_{p}},\quad t_{p}=\frac{1}{\sqrt{\gamma -\alpha }},\quad D=-\frac{\beta }{2\eta ^{2}},\quad \theta =\frac{\beta (\gamma -\alpha )}{2\eta ^{2}},\quad \sigma =2\eta ^{2}.
\end{equation}

\noindent This result is very similar to \( 2\pi  \)-solution (\ref{e7}), however
there is some difference: as the amplitude of the \( \pi  \)-pulse is two times
smaller than the amplitude of \( 2\pi  \)-pulse, the KLM-parameter should
be increased four times and the dispersion should be decreased accordingly in
order to produce the same effect and support coherent pulse generation.

Curve 4 in Fig. \ref{f2} depicts the duration of \( \pi  \)-pulse. As is seen, the
pulse duration slightly differs from the duration of \( 2\pi  \)-pulse (curve
2) and is shorter in the region of small \( \eta  \) (curve 2 in Fig. \ref{f4}).

As was said before, the \( \pi  \)-pulse inverts the population difference
in the absorber that causes the noise amplification behind the pulse tail. Hence,
the laser pulse stabilization is possible if the condition \( \alpha +\gamma _{a}-\gamma <0 \)
is satisfied. This defines the maximal initial loss in the absorber (dotted
curve in Fig. \ref{f3}). It is seen, that the maximal modulation depth exceeds the
threshold (dashed curve) and, consequently, the generation of the stable \( \pi  \)-pulse
is possible.

When the condition \( \eta ^{2}<\frac{3\sqrt{\sigma ^{2}+\beta ^{2}}-\sigma }{4} \)
holds there are the chirped pulses with \textit{sech}-shape. In this
case, the expressions for \( D \) and pulse parameters are bulky and we do
not write them here \cite{My5}. The pulse duration is presented in Fig. \ref{f2} by curve
5. As is seen, the duration of chirped pulse can be very short even for the
moderate value of \( P \). There is a minimum in the dependence of the pulse
duration on \( \eta  \) (an optimal reflectivity of semiconductor absorber
device, curve 3 in Fig. \ref{f4}) and it does not coincide with the point of precise
chirp compensation (curve 1 in Fig. \ref{f5}, a). Additionally, the pulse area is variable
for this type of solution (curve 1 in Fig. \ref{f5}, b).

There is a sharp minimum in the dependence of the pulse duration on \( \sigma  \)
(curve 4 in Fig. \ref{f4}). In our case a corresponding KLM-parameter is \( 7\cdot 10^{7}\; W \).
As it was in the previous case, the minimum of the pulse duration does not coincide
with the point of chirp compensation (curve 2 in Fig. \ref{f5}, a). Unlike the case
of variation of \( \eta  \), the change of \( \sigma  \) causes only a slight
variation of the pulse area (curve 2 in Fig. \ref{f5}, b).

Summarizing, the generation of \textit{sech}-shaped \( \pi  \)-pulses and chirped
pulses with variable area is possible in KLM-laser with coherent semiconductor
absorber as a result of definite relation between KLM's and saturable absorber's
contribution. A larger KLM contribution (for a fixed \( \eta  \)) is needed
to produce the pulse in the comparison with the case of \( 2\pi  \)-pulse generation.

\section{Self-starting ability}

Our results suggest that the pulse duration is determined rather by KLM, whereas
a saturable absorber puts a limitation on the pulse area and stabilizes the
pulse against automodulations. But another important feature of KLM in the presence
of semiconductor absorber is the self-starting ability. To estimate it in our
model we analyzed an evolution of the initial noise spike, which is much longer
than the relaxation time of the excitation in absorber \( T_{a}=1\; ps \).
For such noise spike an absorber is fast and the action of SPM and self-focusing
is negligible. Using the normalization of the time, gain saturation energy and
field intensity to \( T_{cav} \), \( E_{a} \) and \( E_{a}/T_{cav} \), respectively,
an evolution equation for the pulse seed is:

\begin{gather}
\label{e10}
\frac{\partial a(z,t)}{\partial z}=
\left( \frac{P\alpha _{max}T_{r}}{1+\frac{2\tau T_{r}a_{0}^{2}(z)t_{p}(z)}{\eta ^{2}}+PT_{r}}-\frac{\gamma _{a}}{1+2a_{0}(z)t_{p}T_{a}}-\gamma +t_{f}^{2}\frac{\partial ^{2}}{\partial t^{2}}\right) a(z,t),
\end{gather}

\medskip{}
\noindent where all notations have the meaning as before, and the field parameters
refer to the noise spike. To solve Eq. (\ref{e10}) we used, as before, an aberrationless
approximation. The decay of the field (growth of the pulse duration and decrease
of its intensity) means in our model that the system will not self-start. An
opposite situation with an asymptotic growth of the pulse seed testifies about
ability of the system to self-start.

Fig. \ref{f6} demonstrates the regions of the initial pulse parameters corresponding
to the self-starting. The dark zone corresponds to the pump \( P=8.5\times 10^{-4} \),
which is close to the threshold of mode locking self-starting. Lower pump can
not provide the self-starting while a higher pump (\( P=8.8\times 10^{-4} \))
causes an expansion of self-starting region.

\section{Conclusion}

In conclusion, we investigated analytically the conditions of the coherent pulse
formation in cw solid-state laser with the semiconductor absorber. It was found,
that the mode locking in the absence of Kerr-lens-induced fast saturable absorption
does not produce \textit{sech}-shaped pulse. However, there exists \( 2\pi  \)-pulse
(pulse of self-induced transparency), which has fs-duration and is stabilized
by the defined minimal modulation depth of absorber. The stabilization results
from the coherent interaction with absorber in the condition of the positive
difference between saturated gain and linear loss coefficients. But positive
value of this difference increases the threshold of sub-10 fs pulse generation
due to growth of the modulation depth of absorber, that is required by stability
condition. A combined action of KLM and coherent absorption produces the \textit{sech}-shaped
pulse. In this case there is no a requirement to the minimal modulation depth
of semiconductor absorber. The pulse duration, which is close to the fundamental
limit, is defined by KLM and the coherent absorber defines the pulse area, stabilizes
the pulse against the automodulations and self-starts the mode locking operation.
As result, there are the generation of \( 2\pi  \)-, \( \pi  \)-pulses and
chirped pulses with variable area.

Our results can be useful for the development of high-efficient self-starting
generators of extremely short pulses for fs-spectroscopy, X-ray and THz generation.

All calculations in this paper were carried out in computer algebra system Maple V (r5.0) and Maple 6,
the corresponding commented programs are presented on \textit{http://www.\ geocities.com/optomaplev}
and summarized in \cite{My3}.

\vfill\newpage
\section{Figure captions}

\begin{figure}[h]
\caption{\label{f1}} \vspace*{4 cm}
\centering\includegraphics{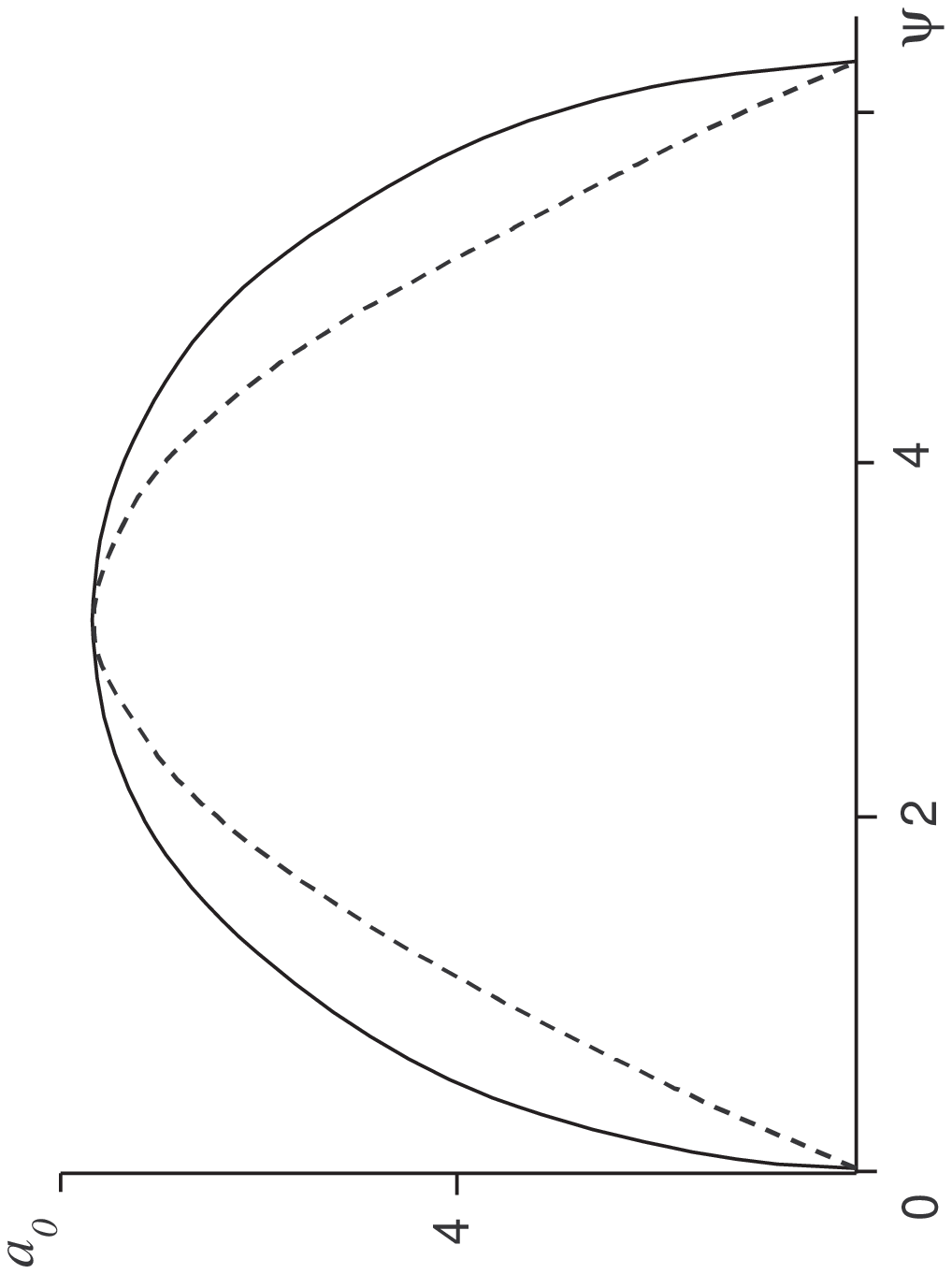}
\end{figure}

Fig. \ref{f1}. \( 2\pi  \)-pulse envelope in the coordinates ``pulse area --
pulse amplitude`` as resulted from numerical solution of Eq. (5) (solid curve)
and the \textit{sech}-shaped pulse envelope (dashed curve). \( \gamma =0.04 \),
\( \gamma _{a}=0.01 \), \( \delta =0.042 \), \( t_{f}=2.5fs \).

\begin{figure}[h]
\caption{\label{f2}} \vspace*{4 cm}
\centering\includegraphics{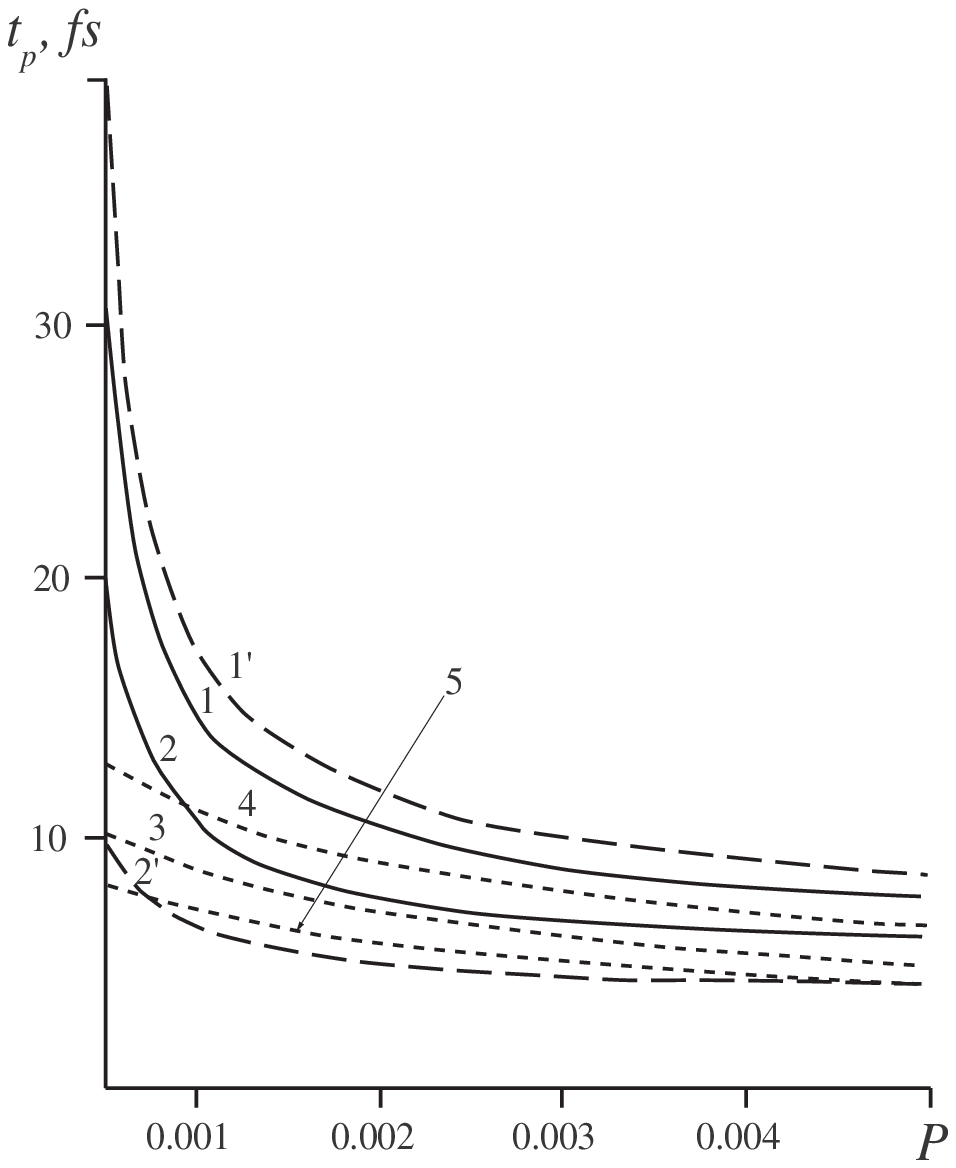}
\end{figure}

Fig. \ref{f2}. Pulse duration \( t_{p} \) versus pump \( P \). \( 2\pi  \)-pulses
(1, 2) in the absence and (1', 2') in the presence of SPM. (2, 2') -- unstable
against the noise solutions. (3) \( 2\pi  \)-pulse; (4) \( \pi  \)-pulse;
and (5) chirped pulse (\( \sigma =0.14 \), \( \beta =0.26 \)) in the
presence of KLM. \( \alpha _{\max }=0.1 \), \( T_{r}=3\; \mu s \),
\( T_{cav}=10\; ns \), \( \tau =6.25\times 10^{-4} \), \( \gamma =0.01 \),
\( \eta =1 \) (1, 1', 2, 2'), \( 0.5 \) (3), \( 0.2 \) (4), \( 0.3 \) (5).

\begin{figure}[h]
\caption{\label{f3}} \vspace*{4 cm}
\centering\includegraphics{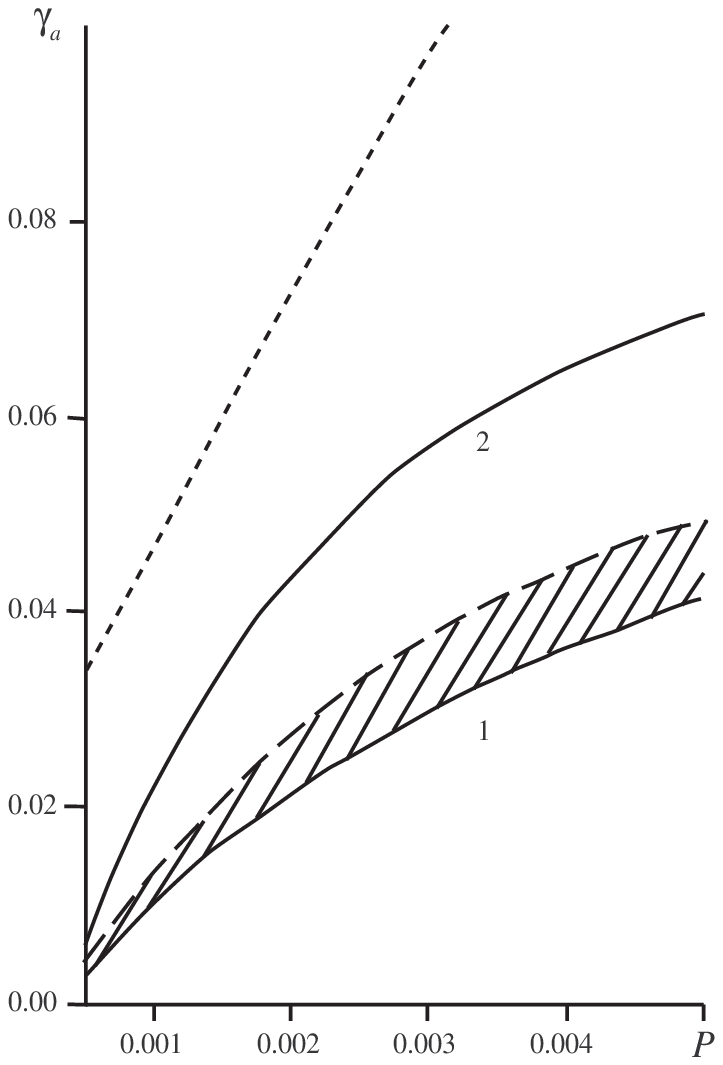}
\end{figure}

Fig. \ref{f3}. Minimal modulation depth of absorber \( \gamma _{a} \) for (1) stable
and (2) unstable \( 2\pi  \)-pulse, (dashed curve) generation threshold and
(dotted curve) maximal modulation depth for \( \pi  \)-pulse. Dashed region
is a stability zone. \( \eta =1 \) (1, 2), \( 0.2 \) (dotted curve).

\begin{figure}[h]
\caption{\label{f4}} \vspace*{4 cm}
\centering\includegraphics{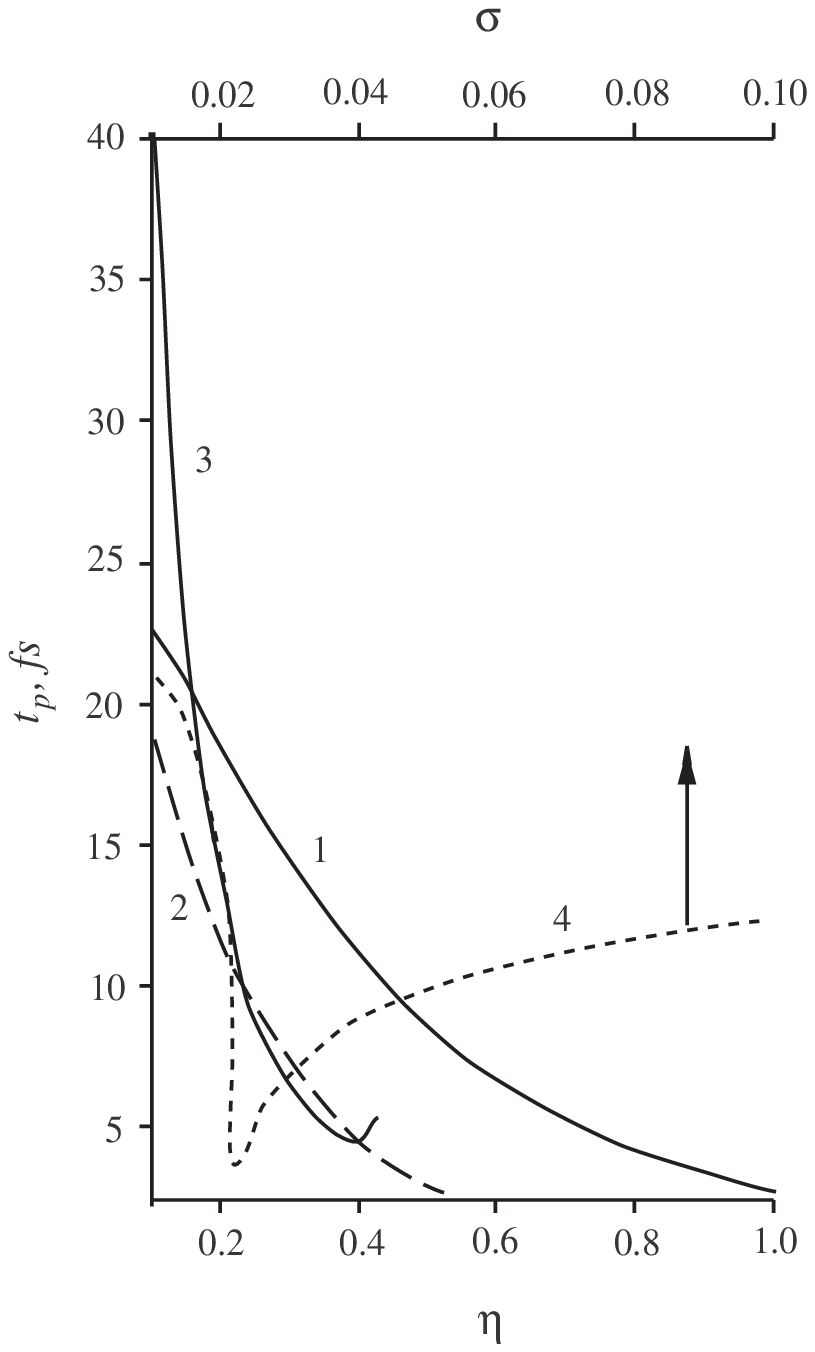}
\end{figure}

Fig. \ref{f4}. Pulse duration \( t_{p} \) versus (solid and dashed curves) \( \eta  \)
and (dotted curve) \( \sigma  \) in the presence of KLM for (1) \( 2\pi  \)-pulse;
(2) \( \pi  \)-pulse; (3) chirped pulse for \( \sigma =0.14 \),
\( \beta =0.26 \); (4) chirped pulse for \( \eta =0.2 \), \( \beta =0.26 \).
\( P=0.001 \) for all curves.

\begin{figure}[h]
\caption{\label{f5}} \vspace*{4 cm}
\centering\includegraphics{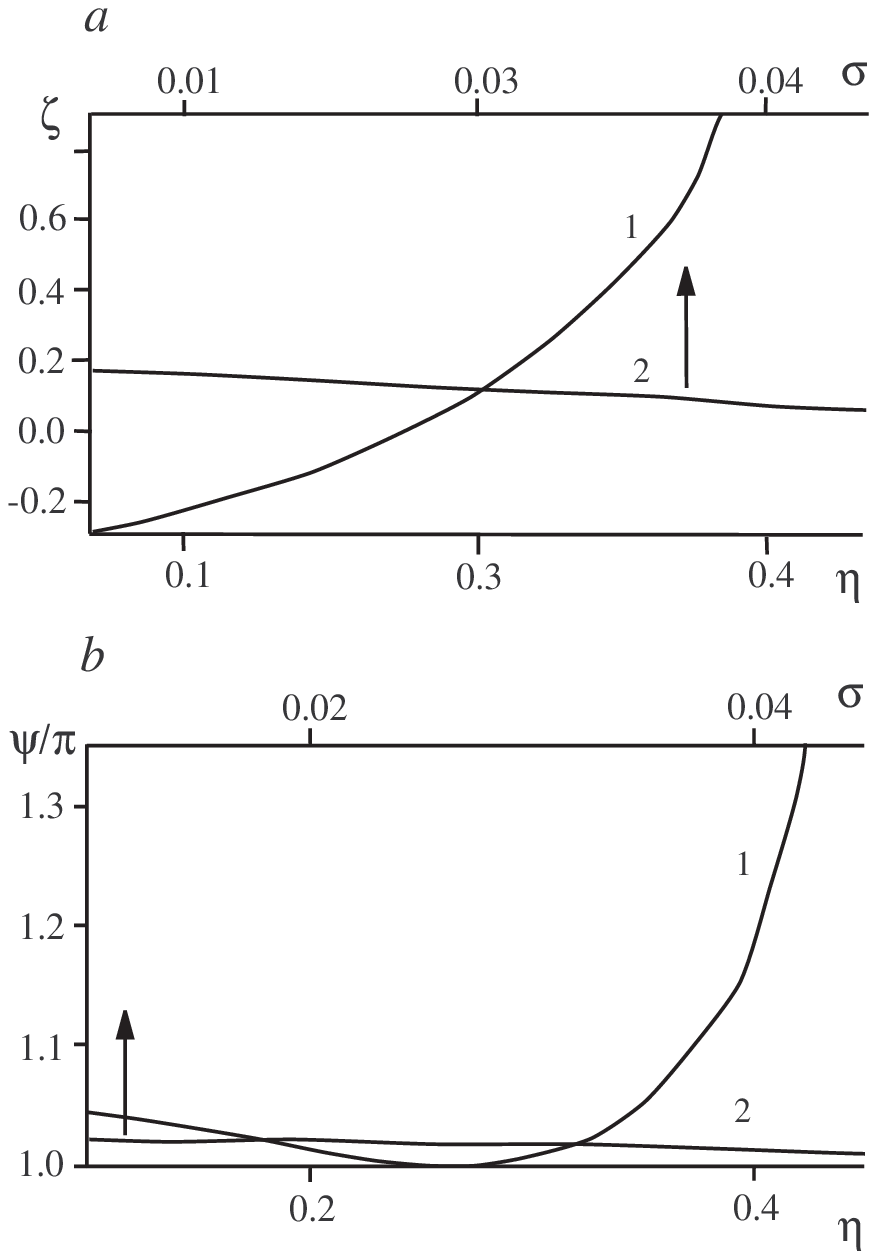}
\end{figure}

Fig. \ref{f5}. a) chirp \( \varsigma  \) and b) pulse area \( \psi /\pi  \) versus
\( \eta  \) and \( \sigma  \) in the presence of KLM: (1) \( \sigma =0.14 \);
(2) \( \eta =0.2 \). \( \beta =0.26 \), \( P=0.001 \).

\begin{figure}[h]
\caption{\label{f6}} \vspace*{4 cm}
\centering\includegraphics{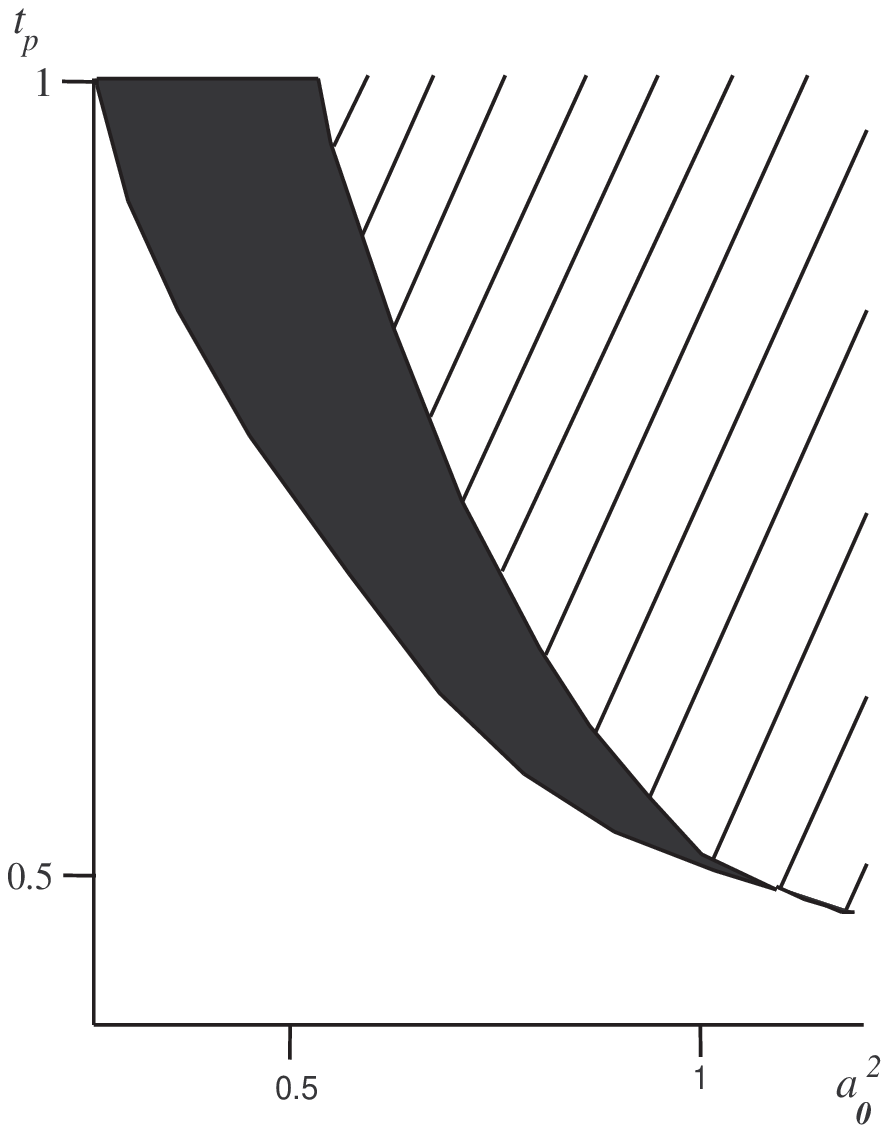}
\end{figure}

Fig. \ref{f6}. Self-starting ranges on the plane ``peak intensity of initial noise
spike -- duration of initial noise spike``. Dark and hatched (together with
dark) regions correspond to the self-starting for \( P=8.5\times 10^{-4} \)
and \( 8.8\times 10^{-4} \), respectively. \( T_{a}=1\; ps \), \( \gamma =0.01 \),
\( \eta =1 \), \( \tau =6.25\times 10^{-5} \).

\vfill\newpage%
\pagestyle{empty}%

\vfill \newpage

\vfill \newpage

\vfill \newpage

\vfill \newpage

\vfill \newpage

\vfill \newpage


\vfill\newpage
\begin{thebibliography}{17}

\bibitem{Keller1} D. H. Sutter, G. Steinmeyer, L. Gallmann, N. Matuschek, F. Morier-Genoud, and U. Keller, Opt. Lett. 24 (1999) 631.

\bibitem{Haus1} H. Haus, J. G. Fujimoto, and E. P. Ippen, IEEE J. Quant. Electr. 28 (1995) 2086.

\bibitem{Keller2} U. Keller, K. J. Weingarten, F. X. K\"{a}rtner, D. Kopf, B. Braun,
I. D. Jung, R. Fluck, C. H\"{o}nninger, N. Matuschek, and J. A. der Au, IEEE J. Selected Topics in Quant. Electr. 2 (1996) 435.

\bibitem{Keller3} F. X. K\"{a}rtner, I. D. Jung, and U. Keller, IEEE J. Selected Topics in Quant. Electr. 2 (1996) 540.

\bibitem{My1} V. L. Kalashnikov, D. O. Krimer, I. G. Poloyko, V. P. Mikhailov, Optics
Commun. 159 (1999) 237.

\bibitem{My2} V. L. Kalashnikov, D. O. Krimer, I. G. Poloyko, V. P. Mikhailov, Proceedings of SPIE, 3683 (1998) 225.

\bibitem{Kozlov1} V. V. Kozlov, E. E. Fradkin, J. Experimental and Theoretical Phys. 80 (1995) 32.

\bibitem{Kozlov2} V. V. Kozlov, J. Experimental and Theoretical Phys. 107 (1995) 360
(in russian).

\bibitem{Komarov} K. P. Komarov, V. D. Ugozhaev, Sov. J. Quantum Electronics, 14 (1984) 787.

\bibitem{Kalosha} V. P. Kalosha, M. M\"{u}ller, and J. Herrmann, J. Opt. Soc. Am. B 16 (1999) 323.

\bibitem{Chr} I. P. Christov, V. D. Stoev, M. M. Murname, H. C. Kapteyn, J. Opt. Soc. Am. B. 15 (1998) 2631.

\bibitem{Allen} L. Allen, J. H. Eberly, Optical resonance and two-level atoms (Wiley,
New York),1975.

\bibitem{Ros} E. N. Rosenwasser, S. K. Volodov, Operator methods and oscillating processes (Nauka, Moskow), 1985 (in russian).

\bibitem{Haus2} H. A. Haus, IEEE J. Quant. Electr. 11 (1975) 736.

\bibitem{My3} V. L. Kalashnikov, e-print arXiv: physics/0009056 (2000).

\bibitem{My4} J. Jasapara, V. L. Kalashnikov, D. O. Krimer, I. G. Poloyko, M. Lenzner, W. Rudolph, J. Opt. Soc. Am. B 17 (2000) 319.

\bibitem{My5} corresponding expressions can be found on \textit{http://www.geocities.com/optomaplev/programs/laser\_pi.html}

\end{thebibliography}
\end{document}